\documentclass[journal=jacsat,manuscript=article]{achemso}
\usepackage{amssymb,amsmath}
\usepackage{graphicx}
\usepackage{float}
\usepackage{bm}
\usepackage{multirow}
\usepackage{dcolumn}
\usepackage{tabularx}
\usepackage{longtable}
\usepackage[dvipsnames,usenames]{color}
\usepackage[latin1]{inputenc} 
\newlength{\dbarheight}

\title{Coupling and Electrical Control of Structural, Orbital and Magnetic
  Orders in Perovskites}
\author{Julien Varignon}
\altaffiliation{These two authors contributed equally}
\affiliation{Physique Th\'eorique des Mat\'eriaux, Universit\'e de Li\`ege
  (B5), B-4000 Li\`ege, Belgium}
\author{Nicholas C. Bristowe}
\altaffiliation{These two authors contributed equally}
\affiliation{Physique Th\'eorique des Mat\'eriaux, Universit\'e de Li\`ege
  (B5), B-4000 Li\`ege, Belgium}
\author{Eric Bousquet}
\affiliation{Physique Th\'eorique des Mat\'eriaux, Universit\'e de Li\`ege
  (B5), B-4000 Li\`ege, Belgium}
\author{Philippe Ghosez}
\affiliation{Physique Th\'eorique des Mat\'eriaux, Universit\'e de Li\`ege
  (B5), B-4000 Li\`ege, Belgium}
\email{philippe.ghosez@ulg.ac.be}

\date{\today}
\begin{document}

\begin{abstract}  
  Perovskite oxides are already widely used in industry and have huge
  potential for novel device applications thanks to the rich physical
  behaviour displayed in these materials. The key to the functional electronic
  properties exhibited by perovskites is often the so-called Jahn-Teller
  distortion. For applications, an electrical control of the Jahn-Teller
  distortions, which is so far out of reach, would therefore be highly
  desirable. Based on universal symmetry arguments, we determine new lattice
  mode couplings that can provide exactly this paradigm, and exemplify the
  effect from first-principles calculations. The proposed mechanism is
  completely general, however for illustrative purposes, we demonstrate the
  concept on vanadium based perovskites where we reveal an unprecedented
  orbital ordering and Jahn-Teller induced ferroelectricity. Thanks to the
  intimate coupling between Jahn-Teller distortions and electronic degrees of
  freedom, the electric field control of Jahn-Teller distortions is of general
  relevance and may find broad interest in various functional devices.
\end{abstract}


\section{Introduction}

Widespread interest in transition metal perovskite-like oxides over the last
several decades can be ascribed to two key discoveries: high-temperature
superconductivity in the cuprates and colossal magnetoresistance in the
manganites.\cite{Possible-LaBaCuO,wu1987Htc-YbaCuO,jonker1950CMR-manganites,jin1994-CMR-manganites}
Physical behavior exhibited by perovskites is by no means limited to these two
phenomena, including ferroelectricity (e.g. titanates) and
(anti)ferro\-magnetism or both simultaneously and coupled in magnetoelectric
multiferroics (e.g. ferrites, manganites), metal-insulator transitions
(e.g. nickelates) and thermoelectricity (e.g. cobaltites), to name a few.  The
wide range of functional properties is usually thanks to an interplay between
the structural (lattice), electronic (orbital and charge) and magnetic (spin)
degrees of freedom allowed within the transition metal
oxides.\cite{zubko2011interface,hwang2012emergent,tokura2000orbital,dagotto2005complexity}
This playground for novel materials physics is not only of fundamental
academic interest, but oxide perovskites have already entered industry and
have huge potential for novel device
applications.\cite{ME-memory,mannhart2010-applicationsME,scott2007applications}

The possibility of tuning the magnetic properties of a material with an
applied electric field has received particular attention for low energy
consumption spintronic devices.\cite{ME-memory,scott2007data} In this regard,
a promising route to achieve ferroelectricity in magnets is the so-called
rotationally driven ferroelectricity.\cite{PTO-STO,NaLaMnWO6,HIF-Rondinelli}
Here one or more antiferrodistortive (AFD) motions, wich are ubiquitous in
perovskites, simultaneously drives the polarization and can couple to the
magnetic orders.\cite{Ca3Mn2O7-Fennie,BFO-LFO,Antiferro-HIF} However, the AFD
motions are rather weakly linked to the electronic properties and hence the
magnetoelectric coupling is likely not the most efficient. It would be
advantageous to replace the AFD motions by another lattice distortion which
couples directly to the electronic properties. Such a motion common in
perovskites is the Jahn-Teller distortion, however there currently does not
exist a clear and universal recipe to control Jahn-Teller distortion with an
external electric field.

In the present work, we provide a pathway to achieve an electric field control
of Jahn-Teller distortions in perovskites through universal symmetry
arguments. Since Jahn-Teller distortions are intimately connected to
electronic degrees of freedom,\cite{Jahn-Teller-effect} such as magnetism,
orbital orderings and metal-insulator phase transitions to name a few, the
proposed mechanism may find broader interest for novel functional devices
outside the field of magnetoelectrics. This mechanism is completely general,
however for illustrative purposes, we demonstrate the concept on the vanadate
perovskites which exhibit a complex structural ground state including
different Jahn-Teller distortions. The effect is further quantified through
first-principles calculation. In AA'V$_2$O$_6$ superlattices, we reveal an
unprecedented orbital ordering and Jahn-Teller induced ferroelectricity and
demonstrate an electric field control of the magnetization.

\section{Bulk A$^{3+}$V$^{3+}$O$_3$}

Whilst the V$^{4+}$ perovskites (e.g. SrVO$_3$) have been studied mainly for
their interesting metallic properties,\cite{SrVO3} the V$^{3+}$ perovskites
are Mott insulators. A$^{3+}$V$^{3+}$O$_3$ compounds have attracted much
attention since the fifties when they were first
synthesized.\cite{1954-vanadate} During this time, many studies began to
determine their magnetic, electronic and structural
properties.\cite{review_vanadates,Bordet-LaVO3,kawano1994YVO3,Miyasaka-diagAVO3,Sage-diagAVO3,LaVO3-magproperties,YVO3-Nature,Tung-LaVO3-neutron,AVO3-reversal-Tung,nakotte1999magneticYVO3,ulrich2003magnetic,Bandgap-YVO3,arima1995optical,noguchi2000synchrotron,AVO3-khaliullin2001spin,YVO-YTiO-comparison,Blake-YVO3,PavariniDMFT}
A central theme at the core of these properties in vanadates is the so-called
Jahn-Teller (JT) distortion. The famous Jahn-Teller theorem claims that a
material with degenerate electronic states will be unstable towards undergoing
a structural distortion lowering its symmetry to remove the electronic
degeneracy.\cite{JT-theorem} In other words, the JT effect is an electronic
instability that can cause a structural and metal-insulator phase
transition. For instance, in the cubic perovskite symmetry, the crystal field
effect splits the $d$ electron levels into a lower lying degenerate three-fold
$t_{2g}$ and a higher lying degenerate two-fold $e_g$ state. Hence in $3d^2$
systems such as the rare-earth vanadates, a Jahn-Teller distortion is required
to split the $t_{2g}$ levels in order to form a Mott insulating
state.~\cite{noteJT}

In the vanadates, two different Jahn-Teller distortions are
observed.\cite{Miyasaka-diagAVO3,Sage-diagAVO3,noguchi2000synchrotron,kawano1994YVO3}
\begin{figure}
\begin{center}
  \resizebox{16cm}{!}{\includegraphics{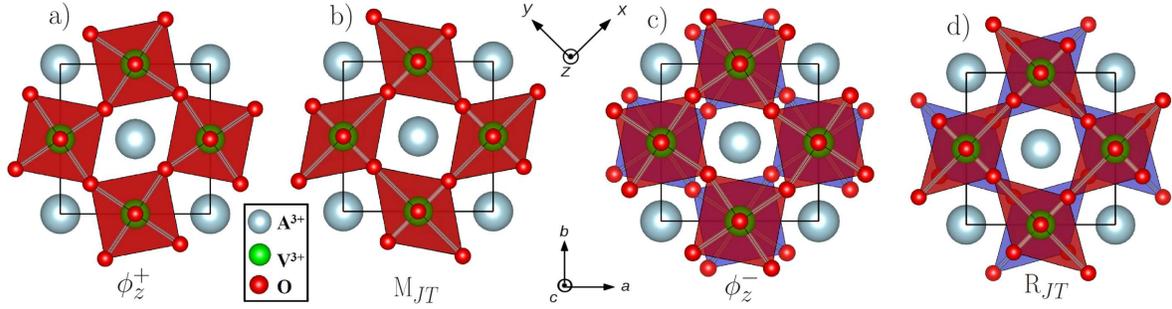}}
\end{center}
\caption{ Comparison of JT and AFD motions around the $z$ axis; a) in-phase
  $\Phi_z^{+}$ AFD motion; b) in-phase M$_{JT}$ Jahn-Teller motion; c)
  anti-phase $\Phi_z^{-}$ motion; d) anti-phase R$_{JT}$ motion. Octahedras
  for the plane in $z$=0 are plotted in red and in blue for the plane in
  $z$=$c$/2. The AFD motions can also appear around the $y$ and $z$ axes (not
  shown), whereas the Jahn-Teller motions only manifest around the $z$ axis in
  the vanadates.}
\label{f:motions}
\end{figure}
The corresponding distortions are displayed in \ref{f:motions} where they are
compared to the so-called antiferrodistortive (AFD) motions that appear
abundantly in perovskites. The AFD motions can be viewed as oxygen octahedra
rotations around an axis going through the B cations, while the JT distortions
correspond to oxygen rotations around an axis going through the A cations.
Both JT and AFD motions can be either in-phase (\ref{f:motions}.a+b) or
anti-phase (\ref{f:motions}.c+d) between consecutive layers and
therefore appear at the M or R points of the Brillouin zone
respectively. Consequently, we label the Jahn-Teller distortions as M$_{JT}$
and R$_{JT}$. While AFD motions do not distort the BO$_6$ octahedra and appear
purely through steric effects, JT motions lift the degeneracy of the $d$
levels through octahedra deformations.  According to such distortions, the
V$^{3+}$ $3d^2$ occupation consists of either a $d_{xy}$ and $d_{xz}$ $t_{2g}$
or a $d_{xy}$ and $d_{yz}$ $t_{2g}$ state.  Nearest-neighbor vanadium sites
within the $(xy)$-plane develop opposite distortions and hence alternative
$d_{xy}$ and $d_{xz}$ / $d_{xy}$ and $d_{yz}$ occupations as shown in the top
panel of \ref{f:oo}.  
\begin{figure}
\begin{center}
  \resizebox{8cm}{!}{\includegraphics{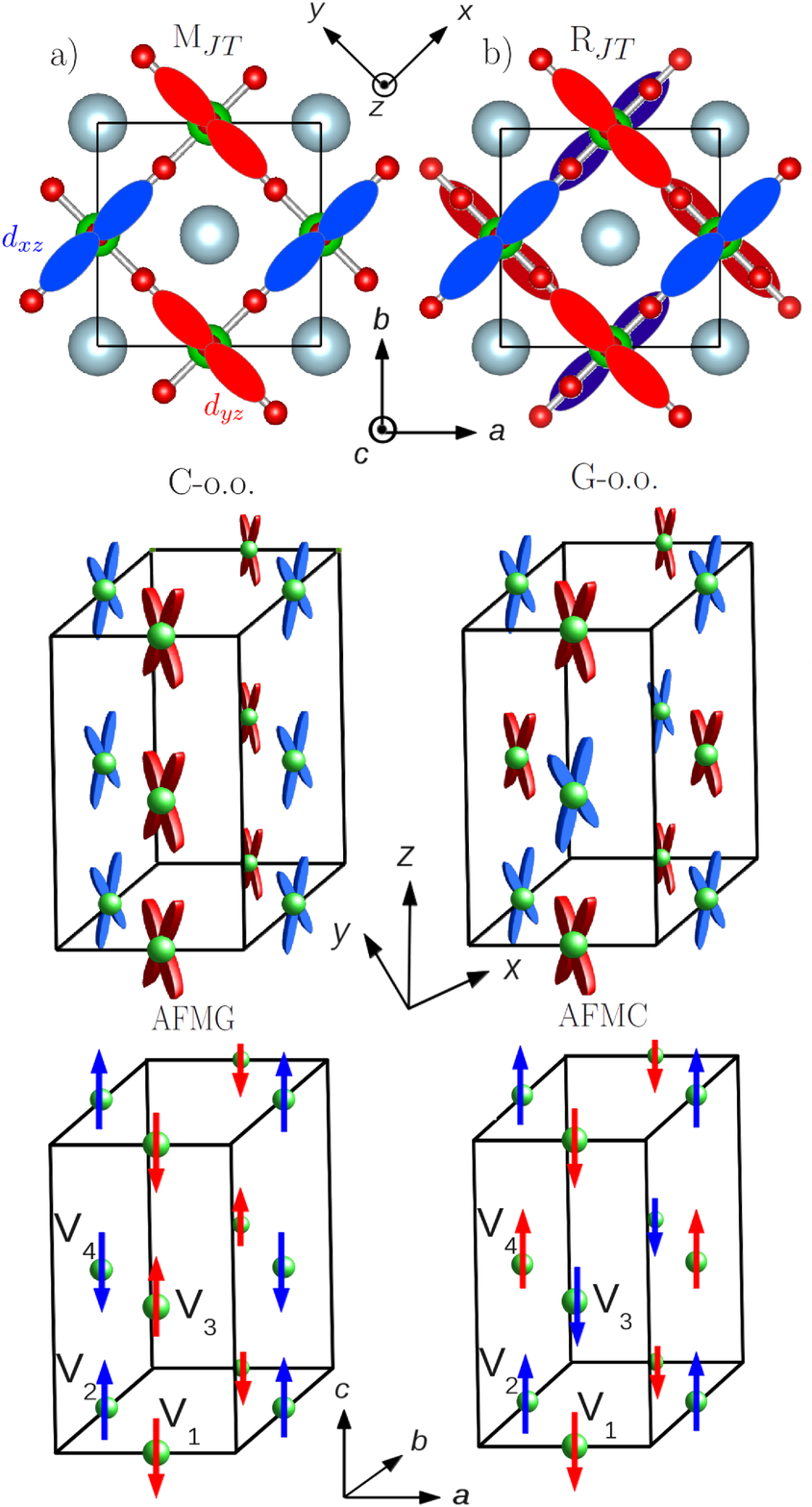}}
\end{center}
\caption{Schematic picture of the idealized orbital orderings between the $d_{xz}$
  (blue) and $d_{yz}$ (red) orbitals. The $d_{xy}$ orbital is not shown for
  clarity. a) Top view of the orbital-ordering induced by a M$_{JT}$
  distortion (top). Consecutive planes exhibit identical orbital orderings
  labelled as a C-type orbital-ordering (C-o.o.) (middle). This results in a
  AFMG ordering (bottom). b) Orbital-ordering induced by a R$_{JT}$ distortion
  (top). Consecutive planes exhibit out-of phase orbital orderings labelled
  as a G-type orbital-ordering (G-o.o.) (middle). This results in a AFMC
  ordering (bottom).}
\label{f:oo}
\end{figure}
Along the $c$ axis, the octahedra deformations
and hence orbital ordering are either in phase (C-type orbital order) or
anti-phase (G-type orbital order) for the M or R Jahn-Teller distortion
respectively (see \ref{f:oo} middle panel). Crucially, the orbital
ordering determines the magnetic ordering through superexchange
interactions.\cite{kramers1934interaction,anderson1950antiferromagnetism,goodenough1955theory,goodenough1958,kanamori1959superexchange}
Strongly overlapping and parallel orbitals between neighboring sites favors
antiferromagnetic superexchange interactions.  With this in mind, the $M_{JT}$
favors a purely antiferromagnetic solution called AFMG whilst the $R_{ JT}$
favors $(xy)$-plane antiferromagnetic alignment and ferromagnetic out-of-plane
alignment called AFMC (see \ref{f:oo} bottom panel). Experiments indeed
observe both AFMG and AFMC magnetic phases in the vanadates, with each
magnetic ordering favoring a certain structural
symmetry.\cite{Miyasaka-diagAVO3,Sage-diagAVO3,noguchi2000synchrotron}

At room temperature, all rare-earth A$^{3+}$V$^{3+}$O$_3$ vanadates
crystallize in a $Pbnm$ structure.  With decreasing temperature, the vanadates
undergo an orbital ordering phase transition to a G-type orbital ordered
(G-o.o.)  phase between 200K and 150K (depending on the A-cation size). This
transition is accompanied by a symmetry lowering from $Pbnm$ to $P2_1/b$. A
magnetic phase transition from a paramagnetic to an AFMC antiferromagnetic
state occurs within this phase at a slightly lower temperature between 150K
and 100K. Finally, for the smallest A cations (A=Yb-Dy, Y), another orbital
ordering phase transition to a purely C-type (C-o.o.) arises and is
accompanied by a structural phase transition from $P2_1/b$ back to $Pbnm$, and
a magnetic phase transition from AFMC to AFMG.  For medium A cations
(A=Tb-Nd), a coexisting orbital-ordering state (C+G-o.o.)  is found with no
structural symmetry change. No further transitions are found for larger A
cations (A=Pr, Ce and La), which remain in the AFMC G-o.o. $P2_1/b$ phase.

It is interesting to note a couple observations. Firstly both the G-o.o. and
AFMC magnetic ordering phases are very rare amongst oxide
perovskites. Secondly the coexisting orbital-orderings appearing for medium
A-size cations is curious, and there is some debate whether this mixed state
is phase separated or coexisting within the same
phase.\cite{Miyasaka-diagAVO3,Sage-diagAVO3} To attempt to shed light on some
of these issues, we perform a symmetry mode analysis of the allowed
distortions with respect to a hypothetical cubic phase on three
different,\cite{Amplimodes1,Amplimodes2} but typical, vanadates: YVO$_3$,
PrVO$_3$ and LaVO$_3$. YVO$_3$ is known to present a $Pbnm$ ground state with
a C-o.o. (AFMG) at low temperature, and a $P2_1/b$ state with a G-o.o. (AFMC)
at higher temperatures, while LaVO$_3$ and PrVO$_3$ only develop a $P2_1/b$
ground state with a G-o.o. character (AFMC). The analysis is performed on
experimental structural data, and the amplitudes of distortions are summarized
in \ref{t:distortions}. The experimental structural data are well
represented in our first principles calculations (see supplementary
material).\cite{Reehuis-YVO3-neutron,Blake-YVO3,Sage-diagAVO3,Seim-LaVO3,Bordet-LaVO3}

\begin{table}
  \centering
\begin{center}
\begin{tabular}{cccccccc}
  \hline & \multicolumn{3}{c}{YVO$_3$} & \multicolumn{2}{c}{PrVO$_3$} & \multicolumn{2}{c}{LaVO$_3$} \\[0.2cm]
  &$Pbnm$ & $P2_1/b$ &$Pbnm$ & $Pbnm$ & $P2_1/b$ & $Pbnm$ & $P2_1/b$ \\ [0.2cm]
  Irreps & 295K\cite{Reehuis-YVO3-neutron} & 100K\cite{Blake-YVO3} &
  5K\cite{Reehuis-YVO3-neutron}& 295K\cite{Sage-diagAVO3} & 5K\cite{Sage-diagAVO3} &   298K\cite{Seim-LaVO3} & 10K\cite{Bordet-LaVO3}\\ [0.2cm]
  \cline{1-8}
  $\phi_{xy}^-$(+$\phi_z^-$)   &-&1.72 & - & - & 1.34& -&1.16\\[0.2cm]
  $\phi_{xy}^-$   &1.71&-&1.73 & 1.34 & - & 1.17 & -  \\[0.2cm]
  $\phi_z^+$  &  1.22	&1.24	&1.22 & 0.94 &0.96 & 0.68 & 0.75     \\[0.2cm]
  M$_{JT}$ & 0.05 &0.06	&0.14 & 0.02 &  0.02 & 0.08 & 0.01 \\[0.2cm]
  R$_{JT}$  & - & 0.10  & - & -& 0.18 & - & 0.08  \\[0.2cm]
  X$_5^{-}$ & 0.86&0.87&0.87& 0.53&0.64& 0.37&0.39 \\[0.2cm]
  X$_3^{-}$ & -& 0.01  & -& -& 0.06 & - & 0.00 \\[0.2cm]
  \hline
\end{tabular}
\end{center}
\caption{Amplitudes of distortions (in \AA ) on experimental structures of vanadates at
  different temperatures. In the $P2_1/b$ symmetry, both $\phi_{xy}^-$ and
  $\phi_z^-$ AFD motions belong to the same irreducible representation. The
  reference structure was chosen as a cubic structure whose lattice vector
  corresponds to the pseudo cubic lattice vector associated to the room
  temperature $Pbnm$ phase. One should notice that for the $P2_1/b$ phase,
  both  $\phi_{xy}^-$ and $\phi_z^-$ motions belong to the same irreps, even
  if the $\phi_z^-$ amplitude should be extremely small.}  
\label{t:distortions}
\end{table}

In the $Pbnm$ phase (a$^-$a$^-$c$^+$ in Glazer's notations)\cite{Glazer}, all
vanadates develop two unique AFD motions $\phi_{xy}^-$
($\phi_{xy}^-$=$\phi_{x}^-$+$\phi_{y}^-$) and $\phi_z^+$ .  Table 1 shows that
the magnitudes of these AFD motions strengthen with decreasing A-cation size
as expected via simple steric arguments.\cite{goldschmidt1926gesetze} In this
symmetry, the M$_{JT}$ distortion is also allowed and amazingly, this
distortion is already present at room temperature and for all A-cations. This
finding is in contradiction with previous reports of a C-o.o. phase only for
small A cations below 100K.\cite{Miyasaka-diagAVO3,Sage-diagAVO3} Our analysis
based solely on experimental data suggests a C-o.o. should already exist at
room temperature, and has likely been overlooked until now. The $Pbnm$ phase
then appears to always be a pure $M_{JT}$ phase. Additionally, an anti-polar
X$_5^-$ mode whose motion is in the $(xy)$-plane is allowed in the $Pbnm$
symmetry. Going to the $P2_1/b$ symmetry, a subgroup of $Pbnm$, the
aforementioned AFD motions are still present, but the R$_{JT}$ distortion is
now allowed and would lead to a G-o.o.  phase. However, in contradiction to
what was sometimes believed,\cite{Miyasaka-diagAVO3,Sage-diagAVO3} the
$P2_1/b$ phase is never an exclusive R$_{JT}$ phase but always coexists with
the M$_{JT}$ distortion, even for the larger A cations (A=Pr, La).  A mixed
C+G-o.o. should then manifest for all $P2_1/b$ structures.  Additionally,
another anti-polar X$_3^-$ mode, whose motion is now along the $z$ direction,
arises in this new phase.

In order to understand the origin and coupling between these distortions, we
can perform a free energy $\mathcal{F}$ expansion (see methods) around a
hypothetical cubic $Pm\bar{3}m$ phase with respect to the different
distortions. In the $Pbnm$ phase, two trilinear couplings are involved in the
free energy:
\begin{eqnarray}
\label{eq:freePbnm}
\mathcal{F}_{tri,\,Pbnm}& \propto & \phi_{xy}^-\, \phi_z^+ \,X_5^-\,\, + \,\, \phi_{xy}^-\,X_5^-\,M_{JT}
\end{eqnarray}
As already mentioned, the two robust AFD modes ($ \phi_{xy}^-$ and $\phi_z^+$)
are the primary lattice motions producing the $Pbnm$ symmetry. When $
\phi_{xy}^-$ and $\phi_z^+$ are non zero in magnitude, the energy of the
system is automatically lowered by the appearance of $X_5^-$ due to the first
trilinear term of Equation \plainref{eq:freePbnm}. Similarly, thanks to the
second trilinear coupling, the energy of the system is lowered by the
subsequent appearance of the $M_{JT}$ distortion. This latter trilinear
coupling explains why the M$_{JT}$ is always present in the $Pbnm$ phase of
vanadates, even at room temperature.  This demonstrates that, in addition to
its possible appearance as an electronic instability, it will always appear as
a structural anharmonic improper mode within the $Pbnm$ phase (whose strength
depends on the coupling constant) even in non magnetic materials. Going to the
$P2_1/b$ phase, two additional trilinear couplings are identified:
\begin{eqnarray}
\label{eq:freeP21c}
\mathcal{F}_{tri,\,P2_1/b} & \propto & \phi_{xy}^-\, \phi_z^+ \,X_5^-\,\, + \,\,
\phi_{xy}^-\,X_5^-\,M_{JT} \nonumber \\
 & + & M_{JT}\, R_{JT}\, X_3^-\, \,  + \phi_z^+\,X_3^-\, \phi_z^-
\end{eqnarray}
As previously mentioned, from experimental observations an orbital-ordering
phase transition to a G-o.o. phase occurs between 150K and 200K for all
vanadates,\cite{Miyasaka-diagAVO3,Sage-diagAVO3} which indicates that the
R$_{JT}$ is likely appearing as a primary electronic
instability. Consequently, through the third trilinear coupling of Equation
\plainref{eq:freeP21c}, both JT distortions produce the additional anti polar
X$_3^-$ motion, in agreement with the experimental data of
\ref{t:distortions}. Finally, an extra $\phi_z^-$ AFD motion arises in the
$P2_1/b$ phase, through the last trilinear coupling.  The combination of
in-phase and anti-phase rotations along the same axis, $\phi_z^-$ and
$\phi_z^+$, in the $P2_1/b$ phase of the vanadates is a rare tilt-system
($a^-a^-c^{\pm}$ in Glazer's notation), previously unknown in bulk perovskites
to the best of the authors' knowledge.

Therefore, within this $P2_1/b$ phase, both JT distortions coexist, but likely
with different origins. The M$_{JT}$ is ``pinned'' into the system through an
improper anharmonic coupling with the robust AFD motions while the R$_{JT}$
appears through the traditional Jahn-Teller electronic instability.  It is
interesting to note that this coexistence is allowed due to the improper
appearance of M$_{JT}$, despite there likely being a competition between both
JTs. This competition would be understood as an electronic origin to favor
one type of orbital ordering over the other, producing a biquadratic coupling
with a positive coefficient in the free energy expansion.  In the light of
there being an abundance of M$_{JT}$ with respect to R$_{JT}$ phases across
the perovskites, we then propose whether it is this improper appearance of
M$_{JT}$ via the robust AFD motions that helps favor this phase universally.
The vanadates would then be a special case where the R$_{JT}$ instability is
robust enough to appear despite this competition.  This universal
  symmetry analysis explains the origin of the coexisting orbital ordered
  phase in the $Pb$ symmetry as observed in vanadates both experimentally and
  from first principles
  calculations.\cite{Miyasaka-diagAVO3,Sage-diagAVO3,PavariniDMFT} 

The coexistence of both JT motions in the vanadates, will also clearly affect
the orbital and magnetic orderings.  One might expect a complex canted
magnetic ordering to occur, resembling partly AFMC and partly AFMG, as
indicated experimentally from neutron scattering on several vanadates. Indeed,
the observed non-collinear spin arrangement in the $P2_1/b$ phase develops an
AFMC ordering in the $(xy)$-plane and a weaker AFMG ordering along the $c$
axis.\cite{Reehuis-YVO3-neutron,Tung-LaVO3-neutron,AVO3-reversal-Tung}
Interestingly, even at the collinear level, our first principles calculations
already indicate this complex magnetic ordering in the $P2_1/b$ phase. Within
the YVO$_3$ $Pbnm$ AFMG C-o.o. phase, all magnetic sites hold roughly the same
magnetic moment ( $1.811 \pm 0.002$ $\mu_B$) indicating a purely AFMG magnetic
ordering as observed experimentally. Going to the $P2_1/b$ AFMC G-o.o. phase
of LaVO$_3$, two magnetic sublattices are observed. Indeed, two different
magnitudes for the magnetic moments are found in consecutive $(xy)$-VO$_2$
layers ($1.822\pm 0.002$ $\mu_B$ and $1.813\pm 0.001$ $\mu_B$) which can be
seen as a dominant AFMC ordering plus a smaller AFMG ordering on the top of
the latter one.

\section{(AVO$_3$)$_1$/(A'VO$_3$)$_1$ layered structures}

Ferroelectric materials possess a spontaneous polarization which is switchable
with an electric field. This electric field controlled multi-state system has
been naturally proposed as an alternative field effect memory device in
electronics.\cite{scott1989ferroelectric} Furthermore, opportunities offered
by magneto-electric multiferroics (electric field control of magnetism and
conversely) would allow for lower energy consumption spintronic
devices.\cite{ME-memory,scott2007data} However, materials combining both
ferroelectric and (anti)-ferromagnetic order parameters are elusive in nature
and the identification of new single phase multiferroics remains a challenge
for modern day research.\cite{d0rule,Revue-novelMF}

In this respect, a common approach to design multiferroics is to engineer
ferroelectricity in magnets.\cite{Revue-novelMF} A standard concept to achieve
this is {\em via} the so-called improper
ferroelectricity.\cite{levanyuk1974improper} Here the polarization is not the
primary order parameter, as in conventional ferroelectrics, but is driven by
one or more primary non-polar modes. A specific emphasis has been dedicated to
the rotationally driven ferroelectricity, where one or more AFD motions induce
the polarization.\cite{PTO-STO} This has enabled ferroelectricity in highly
strained BiFeO$3$,\cite{PhysRevLett.109.057602} in short period ABO$_3$
superlattices,\cite{PTO-STO,HIF-Rondinelli,BFO-LFO,NaLaMnWO6} in
Ruddlesden-Popper systems and even in metal organic
frameworks.\cite{Ca3Mn2O7-Fennie,MOF2-Picozzi}

Following the same spirit, in an attempt to engineer ferroelectricity in
vanadates, we consider (AVO$_3$)$_1$/(A'VO$_3$)$_1$ structures with planes of
different A cations layered along the [001] direction. This structure can
either appear naturally as in the double perovskites, or through single layer
precision epitaxial deposition techniques. The free energy expansion around a
$P_4/mmm$ layered reference structure (equivalent to $Pm\bar3m$ in bulk) then
becomes:
\begin{eqnarray}
\label{eq:freePb}
\mathcal{F}_{tri,\,Pb} & \propto & \phi_{xy}^-\, \phi_z^+ \,P_{xy}\,\, + \,\,
\phi_{xy}^-\,P_{xy}\,M_{JT} \nonumber \\
 & + & M_{JT}\, R_{JT}\, P_z\, \,  + \phi_z^+\,P_z\, \phi_z^-
\end{eqnarray}
The first observation is that the symmetry breaking due to the A cations turns
the X antipolar modes to polar modes, i.e. in-plane (110) $P_{xy}$ and
out-of-plane (001) $P_z$. The first and fourth trilinear couplings of Equation
\plainref{eq:freePb} correspond to the rotationally driven hybrid improper
ferroelectricity mechanism.\cite{PTO-STO,NaLaMnWO6,HIF-Rondinelli} However, we
identify in vanadate superlattices two new trilinear couplings involving JT
distortions (second and third term of Equation \plainref{eq:freePb}). We
especially highlight the third term of Equation \plainref{eq:freePb},
$M_{JT}\, R_{JT}\, P_z\,$, which directly couples the out-of-plane
polarization $P_z$ to both JT distortions. Since JT distortions are intimately
connected to orbital-orderings and particular magnetic states as described in
the previous section, we can expect to have a direct and strong coupling
between polarization and magnetism from this term.

In the present work, we have performed first-principles calculations in order
to show that (AVO$_3$)$_1$/(A'VO$_3$)$_1$ layered structures indeed develop
both in plane and out-of-plane polarizations. $P_{xy}$ appears as a slave of
rotations and is indirectly linked to magnetism through the modification of
the superexchange path as in the usual rotationally driven
ferroelectrics.$^{\text{\cite{Ca3Mn2O7-Fennie}}}$ On the other hand, $P_z$
appears thanks to an electronic instability manifested as a particular orbital
and magnetic ordering.  Finally, we demonstrate that an electric control of
the magnetic state, is indeed possible, providing a novel paradigm for the
elusive magnetoelectric multiferroics.

In order to test the above hypothesis, we considered two different
superlattices: $\rm (PrVO_3)_1/(LaVO_3)_1$ (PLVO) and $\rm
(YVO_3)_1/(LaVO_3)_1$ (YLVO). First principles geometry relaxations (see
method section) of the superlattices converged to two metastable states. As in
the bulk vanadates, a C-type AFM ordering is found in a $Pb$ structure
(equivalent to the $P2_1/b$ in bulk) while a G-type AFM ordering is found in a
$Pb2_1m$ symmetry (equivalent to $Pbnm$ in bulk). We find that PLVO adopts a
$Pb$ AFMC ground state while YLVO adopts a $Pb2_1m$ AFMG ground state. The
symmetry adapted modes and computed polarizations of all metastable phases are
presented in the supplementary material. As predicted, the $Pb2_1m$ ground
state of YLVO only exhibits a $P_{xy}$ polarization of 7.89 $\mu
C.cm^{-2}$. However, the $Pb$ ground state of PLVO develops both $P_{xy}$ and
$P_z$ polarizations of 2.94 and 0.34 $\mu C.cm^{-2}$ respectively. The $P_z$
contribution indicates, as predicted from the third term of Equation
\plainref{eq:freePb}, a Jahn-Teller induced ferroelectricity. Below we explore
the origin of $P_{xy}$ and $P_z$ in more detail.

As we discussed in the first section, bulk vanadates exhibit a $Pbnm$ phase at
room temperature and hence both superlattices should first go to the
equivalent $Pb2_1m$ intermediate phase. We therefore begin by providing
insight on the driving force yielding the various distortions within this
phase. For this purpose, we condense different amplitudes ${Q}$ of
distortions (see methods) within the metastable $Pb2_1m$ state of the PLVO
superlattice starting from an ideal $P_4/mmm$ structure (for each potential,
see supplementary material).  Four main distortions are then present in this
$Pb2_1m$ phase: $\phi_{xy}^-$, $\phi_z^+$, $M_{JT}$ and $P_{xy}$. As expected,
the two AFD motions are strongly unstable (approximately 1 eV of energy gains
for each) and are the primary order parameters of this $Pb2_1m$
symmetry. $P_{xy}$ and $M_{JT}$ present single wells which are the signature
of an improper anharmonic appearance.$^{\text{\cite{nature-Ph}}}$ Therefore,
the $P_{xy}$ polarization appears through a hybrid improper mechanism driven
by the two rotations through the first term of Equation
\plainref{eq:freePb}. Furthermore, as predicted in the first section, this
analysis suggests that the $M_{JT}$ appears with a structural hybrid improper
mechanism rather than an electronic instability.

Having considered the intermediate $Pb2_1m$ phase, we next turn our attention
to the phase transition of PLVO to its $Pb$ ground state. Curiously, a phonon
calculation on the intermediate $Pb2_1m$ phase did not identify any unstable
modes, indicating that no lattice motions can be responsible for the phase
transition. Clearly, the system has to switch from G to C-type AFM and
therefore in an attempt to understand this phase transition we performed two
sets of calculations. The atomic positions were fixed to the intermediate
$Pb2_1m$ structure and the energy was computed i) with imposed and ii) with no
imposed, $Pb2_1m$ symmetry for the electronic wavefunction, both within the
two possible magnetic states. While for the AFMG calculations, no energy
difference is observed between calculations with and without symmetry, the
AFMC calculation with no symmetry leads to a lower energy (around 4.5 meV)
than the one with imposed symmetry. The only difference between the two
calculations is that the electronic structure is allowed to distort and
consequently break the symmetry. We discover that, even with the atoms fixed
in centrosymmetric positions (along the $z$ axis), the electronic instability
creates an out-of-plane polarization $P_z$ of 0.04 $\rm \mu C.cm^{-2}$.

\begin{figure}
  \centering
 \resizebox{16cm}{!}{\includegraphics{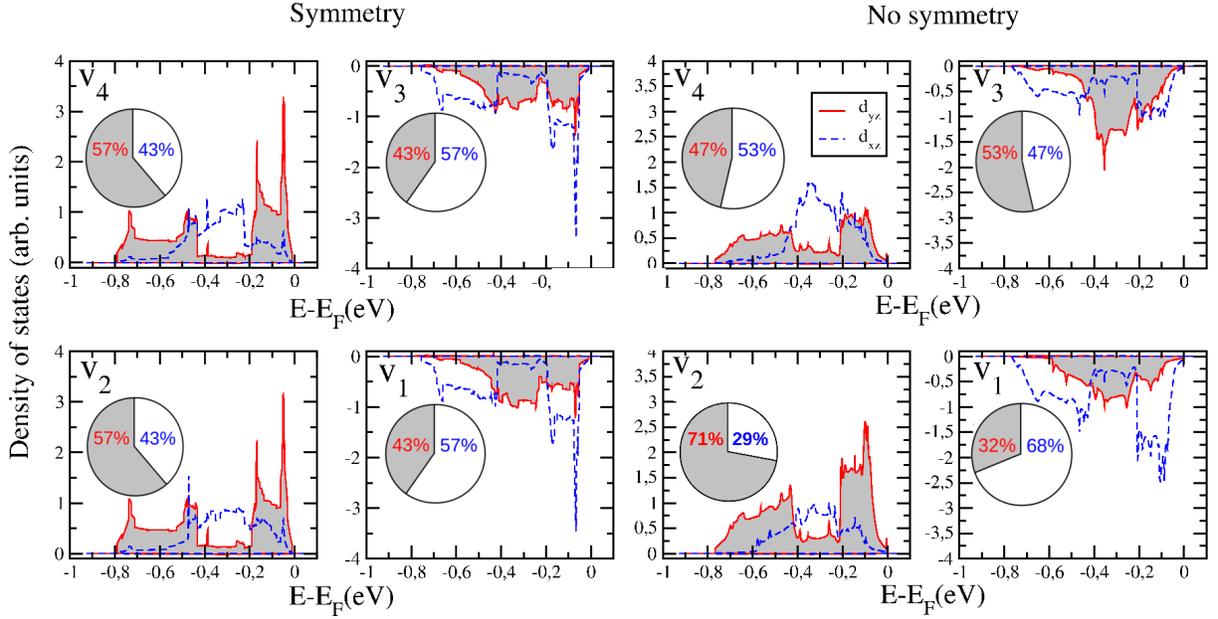}}
 \caption{Projected density of states on the d$_{yz}$ (grey filled red curve)
   and d$_{xz}$ (unfilled blue dashed curve) orbitals of vanadium by imposing
   the $Pb2_1m$ atomic positions and computing the energy with $Pb2_1m$
   symmetry (left) for the electronic wavefunction and by removing the
   symmetry (right) for the electronic wavefunction. $d_{xy}$ is not displayed
   for clarity. V$_1$ (V$_3$) and V$_2$ (V$_4$) are located within the same
   (001) plane as defined in \ref{f:oo}.  The percentage of total
   d$_{yz}$ (red) and d$_{xz}$ (blue) character on each vanadium is shown as a
   pie chart to illustrate the change from C-o.o to C+G-o.o. once the symmetry
   constraint for the wavefunction is lifted.}
\label{f:DOSS}
\end{figure}

In order to understand the nature of this electronic instability, we plot the
projected density of states on vanadiums in \ref{f:DOSS}.
Starting from the projected density of states with $Pb2_1m$ symmetry,
consecutive atoms along the $z$ direction (V$_1$ and V$_3$, V$_2$ and V$_4$ on
\ref{f:oo}) exhibit identical density of states. Consequently, the orbital
ordering appears to be of C-type. When allowing the electronic structure to
distort, several changes appear in the orbital occupations. Consecutive atoms
along the $z$ direction now prefer to occupy either more of the $d_{xz}$ or
the $d_{yz}$ orbital, which results in a mixed G-type plus C-type orbital
ordering. The G-o.o. that appears, despite the absence of the $R_{JT}$ motion,
is allowed {\em via} the Kugel-Khomskii mechanism.\cite{KugelKhomskii} This
mixed orbital ordering produces an asymmetry between the VO$_2$ planes, as
indicated by the two magnitudes of magnetic moments in each layer
(1.816$\pm$0.001 $\mu _B$ and 1.819$\pm$0.001 $\mu _B$). The mixed orbital
ordering also appears in the bulk vanadates, such as the G-o.o.+C-o.o. ground
state of LaVO$_3$ (previously thought to be just G-o.o. from
experiments).\cite{Miyasaka-diagAVO3,Sage-diagAVO3} However, here it is not
enough to break the inversion symmetry along the $z$ axis yielding no
out-of-plane polarization. The second necessary ingredient is the symmetry
breaking due to the A and A' ordering along the [001] direction in the
superlattices. The combination of both effects (in the AO and VO$_2$ planes)
is required to break inversion symmetry along the $z$ axis and to produce the
out-of-plane polarization.  The result is an orbital ordering induced
ferroelectricity in vanadate superlattices.

\begin{figure}
\begin{center}
  \resizebox{16cm}{!}{\includegraphics{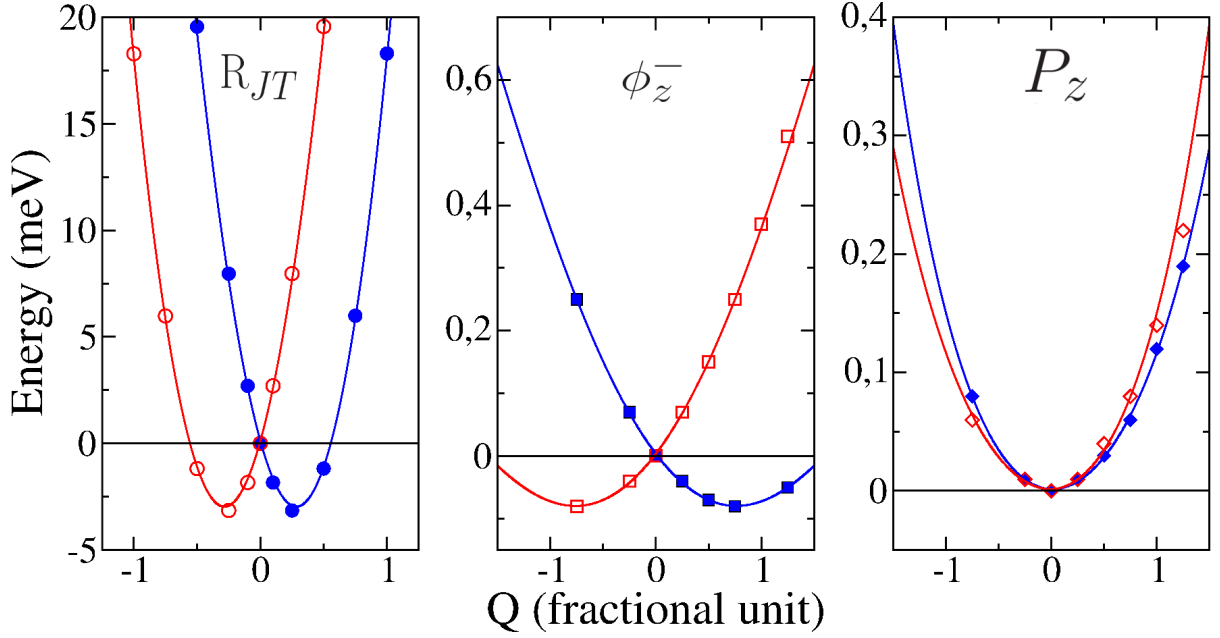}}
\end{center}
\caption{Energy gains (in meV) by condensing various amplitudes of the
  individual $R_{JT}$, $\phi_z^-$ and $P_z$ distortions. Potentials are
  plotted for a '$\uparrow$' (filled blue symbols) and a '$\downarrow$'
  (unfilled red symbols) initial electronic polarization.}
\label{f:potentiel-Pb}
\end{figure}

Interestingly, the direction of the orbital ordering induced ferroelectric
polarization is found to be arbitrary, and both +0.04 and -0.04 $\rm \mu
C.cm^{-2}$ are observed. Each state displays a reversal of the magnitude of
the magnetic moment of the two VO$_2$ planes.  Starting from these two
possibilities, we performed the geometry relaxation and it ended with the
previously identified $Pb$ ground states, with both possibilities (up and
down) for the out-of-plane polarization. We note that the difference in
magnetic moment between both VO$_2$ planes is more pronounced (1.820$\pm$0.001
$\mu _B$ and 1.828$\pm$0.001 $\mu _B$) after the geometry relaxation. Three
new lattice distortions develop to reach the $Pb$ phase: $P_z$, $R_{JT}$ and
$\phi_z^-$. To understand the nature of their appearance, we plot in
\ref{f:potentiel-Pb} each potential as a function of its amplitude
${Q}$.
All potentials present single wells, more or less shifted through an improper
coupling with the electronic instability. This confirms that the electronic
instability is the primary order parameter driving the phase
transition. Moreover, the $R_{JT}$ motion presents an energy gain of one to
two orders of magnitude larger than those of the $\phi_z^-$ motion, indicating
that the $R_{JT}$ couples more strongly with the electronic instability, which
might be expected.  Consequently, once the electronic instability condenses,
the $R_{JT}$ lattice distortion is forced into the system and finally, this
latter produces the lattice part of the polarization through the structural
hybrid improper coupling. This Jahn-Teller induced ferroelectricity amplifies
by one order of magnitude the electronic out-of-plane polarization.  The sign
of the three lattice distortions is again imposed by the initial sign (up or
down) of the electronic polarization. Consequently, the reversal of $P_z$
through an application of an external electric field would require the
reversal of both $R_{JT}$, $\phi_z^-$ and the magnitude if the magnetic moment
of both VO$_2$ planes. The saddle point at the midway of this reversal (all
three modes equal zero, i.e. the $Pb2_1m$ phase) is of the order of 10 meV
higher in energy, which represents a reasonable estimate of the ferroelectric
switching barrier. Compared to the rotationally driven ferroelectricity
$P_{xy}$, whose energy barrier is of the order of 0.1 to 1
eV,\cite{Antiferro-HIF,HIF-Rondinelli,BFO-LFO} this Jahn-Teller induced
ferroelectricity is therefore very likely to be switchable. The large
difference between the two energy barriers is due to two different energy
landscapes involving i) the robust AFD motions inducing $P_{xy}$ and ii) the
relatively soft distortions inducing $P_z$.

Finally we discuss a novel route to create the technologically desired
electrical control of magnetization.  Starting from a $Pb2_1m$ phase with an
AFMG magnetic ordering, the application of an external electric field $\vec E$
along $z$ will induce $P_z$ in the system through the dielectric effect. As a
result, the $R_{JT}$ distortion is automatically induced through the
$M_{JT}R_{JT}P_z$ trilinear term. This electric field induced $R_{JT}$
distortion is a general result for any $(ABO_3)_1/(A'BO_3)_1$ superlattice
consisting of two $Pbnm$ perovskites. Since $R_{JT}$ distortions are
intimately connected to the G-o.o.  and the AFMC magnetic ordering, for a
finite value of $\vec E$, the system may switch from the initial AFMG phase to
the AFMC phase. In reality, the AFMC phase should exhibit a net weak
magnetization from a non collinear magnetic structure as observed in several
bulk vanadates of $P2_1/b$ symmetry.\cite{AVO3-reversal-Tung,YVO3-Nature}
Therefore, the application of an electric field may not only switch between
AFM orderings, but also produce a net magnetic moment in the
material. However, even at the collinear level in our calculations, we can
look at the relative stability between the two magnetic states under an
external electric field in the YLVO superlattice, which presents the desired
$Pb2_1m$ ground state (8 meV lower than the $Pb$ phase).

\begin{figure}
\begin{center}
\resizebox{16cm}{!}{\includegraphics{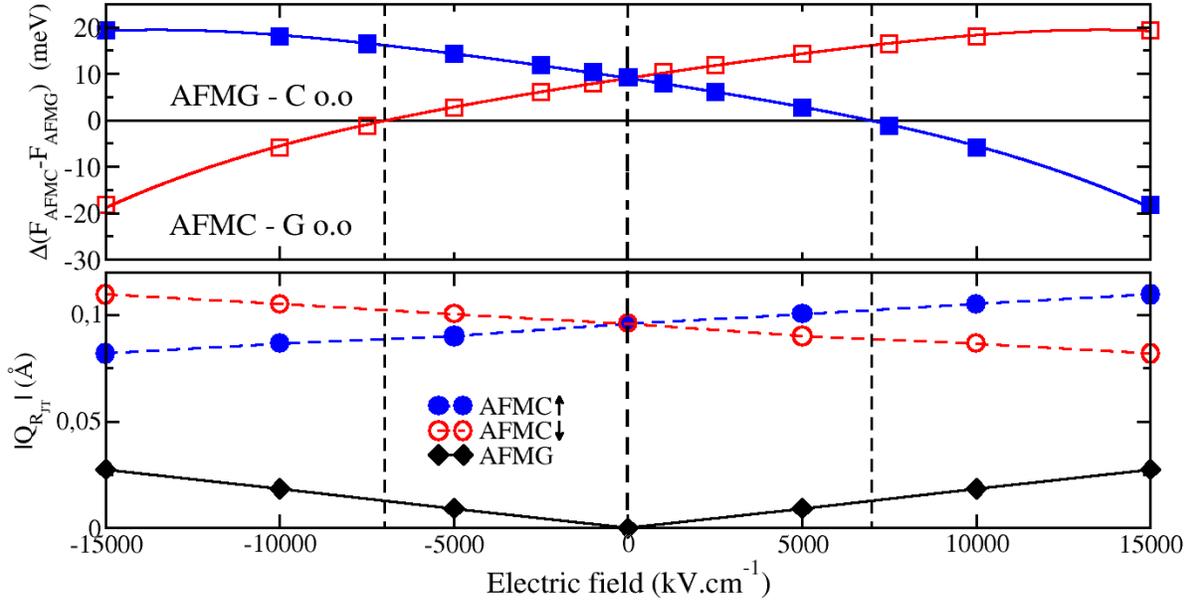}}
\end{center}
\caption{(top) Free electric energy difference (in meV) between the two
  different phases of YLVO superlattices under an applied electric field along
  the $z$ direction. The red unfilled (blue filled) symbols correspond to a
  configuration starting with an $\downarrow$ ($\uparrow$) electronic
  polarization. (bottom) Amplitude of R$_{JT}$ distortion as a
    function of the electric field applied along the $z$ direction in the
    $Pb2_1m$-AFMG phase (filled diamonds) and in the $Pb$-AFMC phase (circles).  }
\label{f:electric}
\end{figure}

\ref{f:electric} (top panel) plots the free electric energy (see
  methods) between the two phases as a function of an electric field applied
  along the $z$ direction.
As the electric field increases in the system, the relative free energy
difference between the two phases reduces, and indeed at around 7
MV.cm$^{-1}$, the $Pb$-AFMC phase becomes more stable than the initial
$Pb2_1m$-AFMG phase. This electric field value corresponds to a voltage of
0.55 V for one bilayer (c$\simeq$7.8~\AA). The critical electric field can be
minimized by reducing the energy difference between the two phases at zero
field.  This can be achieved by for example changing the rare earth cations or
applying biaxial epitaxial strain (see supplementary material).

More generally, we demonstrate in the present study an electric field control
of the Jahn-Teller distortions (see \ref{f:electric} bottom). Since this
mechanism arises from universal symmetry relations, we can expect this effect
to also appear in other perovskite superlattices. This effect may find
applications outside the field of magnetoelectrics such as for tunable band
gaps and metal-insulator transitions, since the Jahn-Teller distortion affects
the electronic structure in general.

\section{Conclusions}
In conclusion, we have identified novel lattice mode couplings in the
vanadates, helping to clarify the origin of the unusual coexisting Jahn-Teller
phase, and indeed the role of Jahn-Teller distortions in perovskites in
general.  These findings have enabled the prediction of a novel paradigm for
the elusive magnetoelectric multiferroics, based on a Jahn-Teller / orbital
ordering induced ferroelectricity.  Due to the intimate connection between
Jahn-Tellers and orbital ordering with magnetism, this unprecedented type of
improper ferroelectric facilitates an electric field control of magnetization.
The rationale is completely general, and a challenge for applications will be
to identify new materials with a magnetic and co-existing Jahn-Teller phase at
room temperature. The demonstration of an electric field control
  of Jahn-Teller distortions may find more general applications for novel
  functional devices, outside the field of multiferroics. We hope these
discoveries will help motivate future studies that will further unlock the
potential of vanadate perovskites, and other Jahn-Teller systems.

\section*{Methods}

The basic mechanism we propose here is solely based on symmetry
arguments. Symmetry mode analysis of experimental data were performed using
amplimodes.\cite{Amplimodes1,Amplimodes2} The free energy expansion of
Equations~\plainref{eq:freePbnm},~\plainref{eq:freeP21c} and
\plainref{eq:freePb} is performed using the invariants software from the
isotropy code.\cite{ISOTROPY} The results from these symmetry considerations
are not dependent on the technical parameters of the first-principles
calculations. The latter are only there to illustrate on a concrete basis and
quantify the effect. First principles density functional theory calculations
were performed using the VASP package.\cite{VASP1,VASP2} We used a
$6\times6\times4$ Monkhorst-Pack k-point mesh to model the $Pbnm$ ($P2_1/b$)
phase and a plane wave cut off of 500 eV.\cite{MonkhorstPack} Optimized
Projector Augmented wave (PAW) potentials for PBEsol exchange-correlation
functional were used in the calculations.\cite{PBEsol} The polarization was
computed using the Berry phase approach as implemented in
VASP.\cite{Resta-1,KSV} The study was performed within the LDA+U framework
using an effective U$_{\rm eff}$.\cite{LDAU,LDAU-Lich,LDAU-Duda} The LDA+U
framework has already been shown to be sufficient to reproduce the ground
state of vanadates.\cite{PavariniDMFT,fang2004quantum} The effective U$_{\rm
  eff}$ parameter was first fitted on bulk compounds in order to correctly
reproduce the ground state of the bulk vanadates. A value of U$_{\rm eff}$=3.5
eV was obtained (see Table 2 and Table
3 
in the supplementary material).  Phonon calculations were performed using the
density functional perturbation theory.\cite{RMP-Baroni} We utilized a
collinear approach to model the magnetic structures. Structural relaxations
were performed until the maximum forces were below 5 $\mu$eV.\AA$^{-1}$ and
the energy difference between conjugate gradient steps was less than $10^{-9}$
eV. The superlattices were relaxed starting from four different initial
guesses: two magnetic orderings (C- and G-type AFM) and two space groups
($Pb2_1m$ and $Pb$, subgroups of $Pbnm$ and $P2_1/b$ respectively for the
layered structures). Lattice distortion potentials were plotted as a function
of the amplitude ${Q}$ of each mode separately.  ${Q}$ is defined as the
fraction of the amplitude of the distortion with respect to the ground state.
The electric field effect on the system was modeled using a linear response
resulting in an ionic relaxation.\cite{Gonze-97} Since an electric field is
applied on the system , the free energy of the system becomes
\begin{equation}
  \mathcal{F} = E_{KS} - \Omega \vec E \cdot \vec P
\label{e:freeelectric}
\end{equation}
where $E_{KS}$ is the Kohn-Sham energy, $\Omega$ is the unit cell volume.

\begin{acknowledgement}
Ph. Ghosez acknowledges Research Professorship from the
Francqui foundation and the ARC project TheMoTherm. Calculations have been
performed within the PRACE projects TheoMoMuLaM and TheDeNoMo and on Nic4
cluster at ULg.  Authors acknowledge fruitful discussions with D. Fontaine,
J.~M. Triscone and
M. Verstraete.
\end{acknowledgement}


\section{Supplementary Material}
\subsection{Optimized bulk vanadates at 0K}

In order to extract the effective U$_{\rm eff}$ parameter for our DFT
calculations, we fitted its value on bulk vanadates in order to correctly
reproduce the ground state of YVO$_3$, PrVO$_3$ and LaVO$_3$. With an
effective parameter of U$_{\rm eff}$ of 3.5~eV, we correctly reproduce the
$Pbnm$ ground state of YVO$_3$ and the $P2_1/b$ ground state of both PrVO$_3$
and LaVO$_3$. Optimized lattice parameters and symmetry mode analysis are
given in \ref{t:bulkParam} and~\ref{t:distBulkDFT} respectively. Within
our DFT calculations, YVO$_3$ exhibits a gap of 1.95 eV in close comparison
with experimental value (around 1.6 eV),\cite{Bandgap-YVO3} while LaVO$_3$ and
PrVO$_3$ develop band gaps of 1.70 eV and 1.93 eV respectively.
\begin{table}
\begin{tabular}{cccccc}
  \hline
  & & \multicolumn{2}{c}{YVO$_3$} & PrVO$_3$ & LaVO$_3$ \\
  & & $Pbnm$ & $P2_1/b$ & $P2_1/b$ & $P2_1/b$ \\
  & & AFMG & AFMC & AFMC & AFMC\\[0.1cm]
  \hline
  \multirow{2}{*}{a(\AA)} & th.  & 5.28 & 5.27 & 5.48 & 5.55 \\
  & exp. & 5.29(5K)\cite{Reehuis-YVO3-neutron} & 5.28(85K)\cite{Reehuis-YVO3-neutron} & 5.48(5K)\cite{Sage-diagAVO3} & 5.56(10K)\cite{Bordet-LaVO3} \\[0.2cm]
  \multirow{2}{*}{b (\AA)} & th.  & 5.61 & 5.65 & 
  5.68 & 5.64\\
  & exp. & 5.59(5K)\cite{Reehuis-YVO3-neutron} & 5.62(85K)\cite{Reehuis-YVO3-neutron} & 5.61(5K)\cite{Sage-diagAVO3} & 5.59(10K)\cite{Bordet-LaVO3}\\[0.2cm]
  \multirow{2}{*}{c (\AA)} & th.  & 7.58 &7.55&   7.72 &7.76\\
  & exp. & 7.56(5K)\cite{Reehuis-YVO3-neutron} & 7.54(85K)\cite{Reehuis-YVO3-neutron} & 7.69(5K)\cite{Sage-diagAVO3}  & 7.75(10K)\cite{Bordet-LaVO3} \\[0.2cm]
  \multirow{2}{*}{$\alpha$ ($^{\circ}$)} & th.  & - &
  90.03 & 90.14 &
  90.15\\
  & exp. & - & 90.02(85K)\cite{Reehuis-YVO3-neutron} &90.15(5K)\cite{Sage-diagAVO3} &  90.13(10K)\cite{Bordet-LaVO3}\\
  \hline
\end{tabular}
\caption{Optimized lattice parameters for the bulk vanadates at 0 K. Experimental
  values are given for comparison.}
\label{t:bulkParam}
\end{table}
\begin{table}
\centering
\begin{tabular}{ccccc}
\hline
 & \multicolumn{2}{c}{YVO$_3$} & PrVO$_3$ & LaVO$_3$ \\
 & $Pbnm$  & $P2_1/b$ & $P2_1/b$ & $P2_1/b$ \\[0.1cm]
\hline
$\phi_{xy}^-$ (+ $\phi_z^-$) &  - & 1.81 & 1.42 & 1.32 \\[0.2cm]
$\phi_{xy}^-$ & 1.83 & -  & - &- \\[0.2cm]
$\phi_{z}^+$ & 1.24 & 1.25 & 0.97 & 0.89 \\[0.2cm]
M$_{JT}$ & 0.15 & 0.06 & 0.02 & 0.01 \\[0.2cm]
R$_{JT}$ & - & 0.10 & 0.10 & 0.09 \\[0.2cm]
X$_5^-$  & 0.90 & 0.90 & 0.65 & 0.52 \\[0.2cm]
X$_3^-$  & - & 0.05 & 0.01 & 0.00 \\[0.2cm]
\hline
\end{tabular}
\caption{Amplitudes of distortions (in \AA) on our optimized bulk
  vanadates at 0 K with respect to a hypothetical pseudocubic phase. In the
  $P2_1/b$ symmetry, both $\phi_{xy}^-$  and $\phi_z^-$ belong to the same
  irreps, even if $\phi_z^-$ amplitude of distortion should remain extremely
  small.}
\label{t:distBulkDFT}
\end{table}

\subsection{Optimized (AVO$_3$)$_1$/(A'VO$_3$)$_1$   layered structures}

\subsubsection{Symmetry mode analysis}
Symmetry mode analysis of our optimized vanadate layered structures are
provided in \ref{t:Dist-SL}.
\begin{table}
\begin{tabular}{ccc}
\hline 
& YLVO (0K) &  PLVO (0K) \\ 
       & $Pb2_1m$-AFMG &  $Pb$-AFMC \\ [0.1 cm]
\hline
$\phi_{xy}^-$ & 1.58 &  1.36 \\ [0.2 cm]
$\phi_z^+$ & 1.13 &0.94 \\ [0.2 cm]
$\phi_z^-$ & - &  0.01 \\ [0.2 cm]
M$_{JT}$ & 0.13 & 0.01 \\ [0.2 cm]
R$_{JT}$ & -  & 0.09 \\ [0.2 cm]
\multirow{2}{*}{P$_{xy}$} & 0.76 & 0.59 \\ 
 & (7.89~$\mu C.cm^{-2}$)  & (2.94~$\mu C.cm^{-2}$) \\[0.2 cm]
\multirow{2}{*}{P$_z$} & - &  0.00(4) \\ 
 & - & (0.34~$\mu C.cm^{-2}$)  \\
\hline
\end{tabular}
\caption{Symmetry allowed distortions (in \AA) with respect to a
  hypothetical $P_4/mmm$ phase of the ground states of the different
  superlattices. A full relaxation of the reference $P_4/mmm$ structures was
  performed before the symmetry mode analysis. Within the $Pb$ symmetry,
  $\phi_{xy}^-$ and $\phi_z^-$ do not belong to the same irreps any more,
  clearly showing a non-zero $\phi_z^-$ contribution, confirming the a$^-$a$^-$c$^{\pm}$ tilt system.}
\label{t:Dist-SL}
\end{table}
The superlattice is insulating and develops a band gap of 1.86 eV.

\subsubsection{Origin of P$_{xy}$}

As discussed in the main paper, bulk vanadates develop a $Pbnm$ symmetry at
room temperature and hence the layered structures should present the equivalent $Pb2_1m$ symmetry.  In the $Pb2_1m$
phase, four main distortions are present: $\phi_{xy}^-$, $\phi_z^+$, M$_{JT}$
and P$_{xy}$. In order to understand the driving force leading to this
metastable $Pb2_1m$ symmetry, we added some amplitude ${Q}$ of
distortions leading to this symmetry in a hypothetical $P4/mmm$
phase. Potentials are plotted in figure~6.
\begin{figure}
 \resizebox{16cm}{!}{\includegraphics{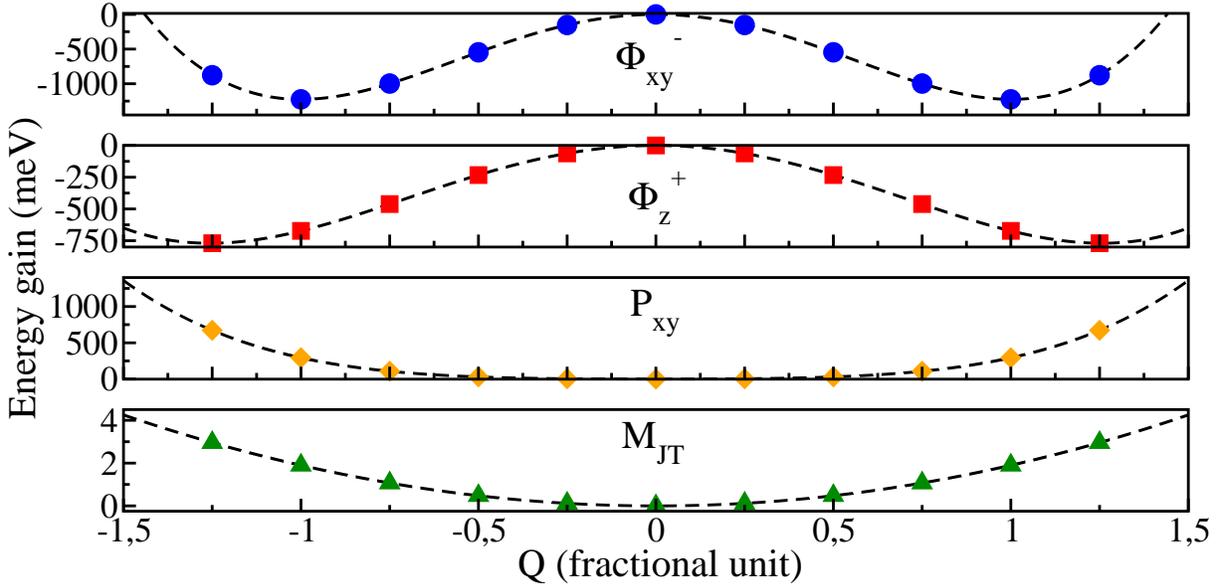}}
 \caption{Energy gains (in meV) by condensing several amplitudes of
   distortions leading to the $Pb2_1m$ (AFMG) within the $P_4/mmm$ structure.}
\label{f:PotP4mmm}
\end{figure}
As expected, the two AFD motions present double wells and are strongly
unstable (energy gains around 1~eV) while P$_{xy}$ and M$_{JT}$ are
stable. Consequently, $\phi_{xy}^-$ and $\phi_z^+$ are the primary order
parameters leading to the $Pb2_1m$ symmetry, and P$_{xy}$ and M$_{JT}$ appear
through improper couplings. These potentials confirm the hybrid improper
character of P$_{xy}$.\cite{PTO-STO,HIF-Rondinelli,NaLaMnWO6}

\subsection{Strain effect on (AVO$_3$)$_1$/(A'VO$_3$)$_1$ layered structures}
For potential applications, the proposed superlattices can be grown on given
substrates. The pseudocubic parameters (defined as the mean value of the low
symmetry lattice parameters) are 3.88~\AA, 3.88~\AA and 3.92~\AA for
YLVO and PLVO superlattices respectively.  We chose cubic SrTiO$_3$
(3.905~\AA) and KTaO$_3$ (3.9885~\AA) as two potential
substrates and performed a geometry relaxation of the
two superlattices constraining both in-plane pseudocubic lattice parameters to
those of the substrates.\cite{schlom2007strain} Results are displayed in \ref{t:toto}.
\begin{table}
\begin{tabular}{cccccccc}
  \hline 
  & &  \multicolumn{2}{c}{SrTiO$_3$} & \multicolumn{2}{c}{Fully relaxed} & 
  \multicolumn{2}{c}{KTaO$_3$}\\
  &  & AFMG & AFMC & AFMG & AFMC & AFMG & AFMC \\[0.1cm]
  \hline
  \multirow{2}{*}{YLVO} & $\Delta E$ & 0 & +24 & 0 & +8 & 0 & -10  \\
  & c/a  & 0.98 & 0.98 & 1.01 & 1.01 & 0.95 & 0.95  \\[0.2cm]
  \multirow{2}{*}{PLVO} & $\Delta E$  &  0 & +8 & 0 & -4 & 0 & -13 \\
  & c/a  & 1.00 & 1.00 & 1.00 & 1.00 & 0.96 & 0.96 \\
  \hline
\end{tabular}
\caption{Energy differences $\Delta E$(in meV) between the two magnetic states  with   repect to the AFMG  structure  on the two different substrates . The fully  relaxed energy difference is given for comparison.}
\label{t:toto}
\end{table}
The effect of the substrates is to tune the relative energy difference between
the two magnetic orderings, and hence between the two orbital-ordered
phases. A SrTiO$_3$ substrate seems to favor an AFMG magnetic ordering, and
consequently a pure M$_{JT}$ phase. On the contrary, going to a KTaO$_3$
substrate applies a moderate tensile strain and favors an AFMC magnetic
ordering for the two superlattices. These strain effects then open the way to
an indirect control of the magnetism and the orbital ordering playing with the
strain imposed by the substrates. Indeed, growing (AVO$_3$)$_1$/(A'VO$_3$)$_1$
layered structures on a piezoelectric substrates, an indirect coupling with
the electric field may appear:
\begin{eqnarray*}
  \text{coupling} & \propto & \frac{electric}{strain} \times  \frac{strain}{magnetism}
\label{eq:indirectcoupling}
\end{eqnarray*} 
Applying an external field on the substrate induces a strain on the vanadate
layered structure, and consequently this latter can switch both magnetic and
orbital orderings.

\end{document}